\documentclass[12pt]{article}
\usepackage{cite,epsfig,amssymb,amsmath,graphicx,color}
\newcommand{\be}{\begin{equation}}
\newcommand{\ee}{\end{equation}}
\newcommand{\bear}{\begin{eqnarray}}
\newcommand{\eear}{\end{eqnarray}}
\newcommand{\ba}{\begin{array}}
\newcommand{\ea}{\end{array}}
\newcommand{\nn}{\nonumber}

\begin{document}

\begin{center}
{{{\Large \bf  Scalar Perturbation in Symmetric Lee-Wick Bouncing Universe}
}\\[17mm]
Inyong Cho$^{1}$ and O-Kab Kwon$^{2}$  \\[3mm]
{\it $^{1}$Institute of Convergence Fundamental Studies \& School of Liberal Arts, \\
Seoul National University of Science and Technology,
Seoul 139-743, Korea}\\[2mm]
{\it $^{2}$Department of Physics,~BK21 Physics Research Division,
~Institute of Basic Science,\\
Sungkyunkwan University, Suwon 440-746, Korea},\\[2mm]}
{\tt iycho@seoultech.ac.kr,~okab@skku.edu}
\end{center}

\vspace{10mm}

\begin{abstract}
We investigate the scalar perturbation in the Lee-Wick bouncing universe
driven by an ordinary scalar field plus a ghost field.
We consider only a symmetric evolution of the universe and the scalar fields
about the bouncing point.
The gauge invariant Sasaki-Mukhanov variable is numerically solved
in the spatially flat gauge.
We find a new form of the initial perturbation growing during the contracting phase.
After the bouncing, this growing mode stabilizes to a constant mode which is
responsible for the late-time power spectrum.
\end{abstract}
\newpage

\tableofcontents

%\clearpage
\section{Introduction}
%\subsection{Generalized Lee-Wick formalism and Lee-Wick bouncing universe}
Lee and Wick (LW) constructed quantum electrodynamics with the
higher derivative (HD) propagator for a photon~\cite{Lee:1969fy}.
The HD term in the denominator of the propagator improves
the ultraviolet convergence of the Feynman diagram since the propagator
falls off more quickly as momentum grows. A minimal set of the HD term
leads to an additional physical pole in the propagator, corresponding
to a massive LW-photon with the wrong-sign residue.
It seems that the wrong sign gives rise to the instability or the violation of
unitarity in the theory. Lee and Wick proposed a deformation of integration
contours in the Feynman diagram so that the theory can be free from the instability
and the violation of unitarity.

Recently
Lee-Wick Standard Model (LWSM) has been proposed as a candidate
to solve the hierarchy problem in the Standard Model~\cite{Grinstein:2007mp}.
Every field in the LWSM has a higher derivative kinetic term.
Using the auxiliary field (AF) method,
the HD term can be converted into the degree of freedom of a massive field
with wrong-sign kinetic term, referred as the LW partner. Since the
LW partner can decay into ordinary fields, the {\it wrong sign} does not
cause the violation of unitarity at macroscopic
scales~\cite{Lee:1969fy,Grinstein:2007iz,Grinstein:2008bg}.
Soon after, the $N=3$ HD theory was constructed by
mapping among the HD Lagrangian, the AF Lagrangian,
and the LW form in Ref.~\cite{Carone:2008iw},
and the generalized $N$ formalism for the scalar field
was constructed in Ref.~\cite{Cho:2010hj}.

Very recently, Cai {\it et. al.} applied the $N=2$ scalar-field Lee-Wick model
to cosmology, and investigated the evolution of the universe~\cite{Cai:2008qw}.
(See also Refs.~\cite{Cai:2009fn}.)
They found that the universe naturally bounces from the contracting phase
to the expanding phase owing to the energy-condition violating
Lee-Wick partner field.
The merit of the bouncing universe in general is that cosmological problems such as
the flatness and the horizon problems are naturally solved~\cite{Cai:2008qw}:
during the contracting phase,
the deviation of $\Omega_K$ from $0$ decreases (solution to the flatness problem),
and the universe starts well inside the horizon (solution to the horizon problem).
The remaining is whether or not, the bouncing universe can provide
proper density perturbations for structure formation.
What the authors found in the Lee-Wick bouncing model are that
the density perturbation at the bouncing point is nonsingular
and the power spectrum can be scale-invariant. See also Refs.~\cite{Zhang:2010bb} 
for the recent works about nonsingular bouncing universe models.

In this work, we investigate the scalar perturbation
and study precisely its properties by inspecting the comoving curvature.
We consider a symmetric background universe about the bouncing point
and numerically solve the scalar perturbation.
We find a new type of initial vacuum solution, and discuss its growth and
resulting late-time power spectrum.

The paper is organized as following.
In Sec.~2, the evolution of the background universe is calculated.
In Sec.~3, the scalar perturbation theory is briefly introduced.
In Sec.~4, the gauge invariant Sasaki-Mukhanov variable is calculated.
In Sec.~5, the comoving curvature is evaluated, the initial perturbation
is obtained, and the late-time power spectrum is discussed.
In Sec.~6, we conclude.

\section{Evolution of background universe}

The general higher derivative Lagrangian of a self-interacting real scalar field $\phi$ is given by
\begin{align}\label{HDL}
{\cal L}_{{\rm HD}}^{(N)} =\frac{1}{2}\sum_{n=1}^N (-1)^{n+1}a_n\phi
\Box^n\phi -\frac{1}{2} m^2\phi^2-V(\phi),\qquad (a_1=1),
\end{align}
where $\Box\equiv \frac{1}{\sqrt{-g}}
\partial_\mu\big(\sqrt{-g}g^{\mu\nu}\partial_\nu\big)$,
$N$ denotes the number of physical poles in the propagator
of $\phi$, $a_n$ is a constant coefficient with  mass dimension $[a_n] =2-2n$,
and $V(\phi)$ represents an interaction potential of the scalar field.
Introducing auxiliary fields we can show the Lagrangian
\eqref{HDL} is equivalent to that of $N$ scalar fields with
quadratic kinetic terms, referred as the Lee-Wick Lagrangian, up to the quantum level~\cite{Cho:2010hj},
\begin{align}\label{LWL}
{\cal L}_{{\rm LW}}^{(N)} = \frac12\sum_{n=1}^{N} (-1)^{n+1}
\varphi_n(\Box - m_n^2)\varphi_n - {\cal V}(\varphi_1,\cdots, \varphi_N),
\qquad (m_1^2<m_2^2\cdots <m_N^2),
\end{align}
where ${\cal V}$ represents the interaction potential originated from $V$ in \eqref{HDL}.

\subsection{Symmetric bouncing universe}
We consider the $N=2$ Lee-Wick model with vanishing interaction potential,
which consists of one ordinary massive scalar field $\varphi_1$
and one ghostlike Lee-Wick partner field $\varphi_2$.
The Einstein-Hilbert action with these matter fields is
\begin{align}
S_{{\rm LW}} = \int d^4x \sqrt{-g}\Big[ \frac{R}{16\pi G} + \sum_{n=1}^{N} (-1)^n
\big(\frac12 \partial_\mu\varphi_n\partial^\mu\varphi_n + \frac12
m_n^2\varphi_n^2\big)\Big].
\end{align}
The corresponding energy-momentum tensor is given by
\begin{align}
T^\mu_{~\nu} =\sum_{n=1}^{N}
(-1)^{n+1}\Big[\partial^\mu\varphi_n\partial_\nu\varphi_n
-\delta^\mu_{~\nu}\big(\frac12\partial_\lambda\varphi_n\partial^\lambda\varphi_n
+\frac12 m_n^2\varphi_n^2\big)\Big].
\end{align}
Nonvanishing components of the energy-momentum tensor for the time-dependent scalar fields are
\begin{align}\label{LWemtensor}
T^0_{~0} &= \sum_{n=1}^{N} (-1)^{n+1}
\Big[-\frac12\dot\varphi_n^2-\frac12m_n^2\varphi_n^2\Big],
\nonumber \\
T^i_{~i} &= \sum_{n=1}^{N} (-1)^{n+1}
\Big[\frac12\dot\varphi_n^2-\frac12m_n^2\varphi_n^2\Big].
\end{align}
With an isotropic metric ansatz describing the background universe,
\begin{align}
ds^2 = -dt^2 + a(t)^2 dx^idx^i,
\end{align}
the Einstein's equation and the scalar field equation read
\begin{align}
&H^2 = \frac{8\pi G}{3}\sum_{n=1}^{N}
(-1)^{n+1}\Big(\frac{1}{2}\dot\varphi_n^2 + \frac12
m_n^2\varphi_n^2\Big),
\label{bgHsq} \\
&\dot H = -4\pi G\sum_{n=1}^{N} (-1)^{n+1} \dot\varphi_n^2,
\label{bgHdot} \\
&\ddot\varphi_n + 3H\dot\varphi_n + m_n^2\varphi_n
= 0. \label{bgmeq}
\end{align}
Since the Lee-Wick partner field is a ghost that violates the null energy condition,
the background evolution can have a bouncing from contracting to expanding.
Let us set the bouncing moment at $t=0$.
In addition, we shall consider a symmetric background about $t=0$.
In order to have a {\it symmetric bouncing} universe,
we need to set conditions at $t=0$.
The choice of those conditions is not unique,
and we adopt one as following:

\begin{enumerate}
  \item One should have $\dot a(0)=0$, i.e., $H(0)=0$,
as a necessary condition for a bouncing universe.
This makes the left-hand side of \eqref{bgHsq} vanish.
One can arrange the right-hand side in several ways to make it vanish.
We adopt nonvanishing values of $\varphi_n(0)$,
then $\dot\varphi_n(0)$ should vanish for $\varphi_n(t)$ to be symmetric,
i.e., $\varphi_n(t)$'s are even functions.
  \item As $\dot\varphi_n(0)=0$, $\dot{H}(0) =0$ from \eqref{bgHdot}.
This implies $\ddot{a}(0)=0$.
  \item Differentiating \eqref{bgHdot}, it is manifest that
  $\dddot{a}(0)=0$, i.e., $\ddot{H}(0)=0$.
  \item The conditions 1-3 provide a bouncing universe.
  However, for a bouncing from contracting to expanding,
  another condition is needed, $\ddddot{a}(0) >0$, i.e., $\dddot{H}(0)>0$.
  In order to satisfy this condition, we shall see later that one must have $|m_1| <|m_2|$.
\end{enumerate}
\noindent As a whole,
$a(t)$ and $\varphi_n(t)$ are even functions,
and $H(t)$ is an odd function.

\subsection{Numerical solutions}
We numerically solve the field equations \eqref{bgHdot} and \eqref{bgmeq}.
We integrate the field equations from $t=0$ to the $t>0$ region.
What is good to do this is that it is numerically convergent and stable.
Then the solutions for the $t<0$ region are obtained simply by {\it symmetry}.
If one integrated from some moment at $t<0$, the numerical instability
would grow up, and it should be very difficult to overcome the bouncing point.
In addition, even if one could obtain smooth regular solutions connected
to the $t>0$ region, it would never be strictly symmetric about the bouncing point.

Applying the symmetric bouncing conditions at $t=0$ discussed in the previous subsection,
once $m_n$'s are fixed, the only condition remaining
for numerical calculation is $\varphi_1 (0)$, or $\varphi_2 (0)$.
The numerical results for the background fields are plotted in Fig.~1.
For numerical calculations in this paper,
we set $m_2=2m_1=10^{-5}m_p \equiv 10M_r$
and $\varphi_1(0)=(m_2/m_1)\varphi_2(0) = m_p = 10^6 M_r$.\footnote{In this paper,
we introduce a mass unit $M_r = 10^{-6}m_p$, where $m_p$ is the Planck mass,
and all the physical quantities
are expressed in this unit.}

At the bouncing point $t=0$, the scalar fields start to decrease
from their maximum values, and then undergo  damped oscillations about $\varphi_n=0$.
In the $t<0$ region, the evolution is exactly opposite; the amplitudes of oscillations
grow and the scalar fields reach their maxima at $t=0$.
The amplitude of $\varphi_1$ is much larger than that of $\varphi_2$
in our set-up.

While the scalar fields roll down from their maxima for $t>0$,
the universe undergoes accelerating expansion, $\ddot a>0$.
Afterwards, the scalar fields start to oscillate and play the role of
pressureless dust which drives the universe to settle down to
the matter-dominated expansion, $a\sim t^{2/3}$.
When $t<0$, the universe contracts initially as  $a\sim (-t)^{2/3}$,
and then undergoes accelerating contraction till the bounce.
%%%%%%%%%%%%%%%%%%%%%%%%%%%%%%%%%%%%%%%%%%%%%%%%%%%%%%%%%%%%%%%
\begin{figure}%[!h]
\centerline{\epsfig{figure=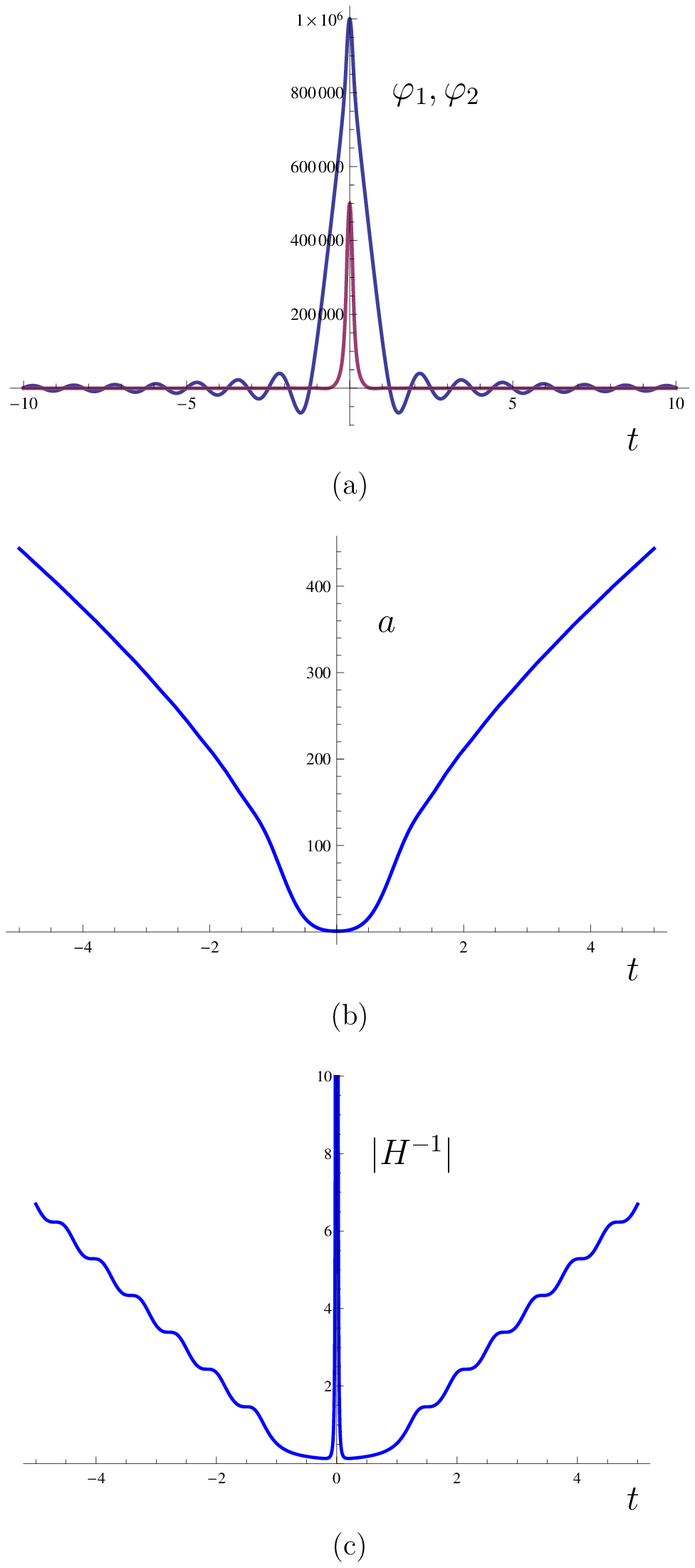,height=180mm}}
\caption{
\small Plot of the background fields which are symmetric about $t=0$ for
$m_2=2m_1=10^{-5}m_p\equiv 10M_r$, $\varphi_1(0)=(m_2/m_1)\varphi_2(0) = m_p=10^6M_r$,
$\dot\varphi_1(0)=\dot\varphi_2(0)=0$, $H(0)=0$, and $a(0)=1$.
(a) Scalar fields $\varphi_1$ and $\varphi_2$ exhibit damped oscillations.
The amplitude of $\varphi_1$ is much larger.
(b) The scale factor $a$ shows the accelerating contraction/expansion
near $t=0$, and $|t|^{2/3}$ contraction/expansion at large $|t|$.
(c) The horizon scale $|H^{-1}|$ diverges near $t=0$ and at large $|t|$.
}
\label{FIG1}
\end{figure}
%%%%%%%%%%%%%%%%%%%%%%%%%%%%%%%%%%%%%%%%%%%%%%%%%%%%%%%%%%%%%%%

The horizon scale $|H^{-1}|$ diverges at $t=0$ and $t\to\pm\infty$.
A moderate comoving scale initially starts well inside the horizon at large $|t|$ $(t<0)$,
and crosses the horizon four times to enter the current universe.

During the matter-dominated period controlled mainly by oscillating $\varphi_1$ at large $|t|$,
the scale factor is approximated by $a\approx \alpha t^{2/3}$
where $\alpha$ is a constant
depending on the values of $m_n$ and $\varphi_n(0)$.
Then field equations \eqref{bgHsq}-\eqref{bgmeq} allow the following
asymptotic solutions,
\begin{align}\label{latephi}
& \varphi_1(t) \approx \frac{\cos (m_1 t + \alpha_1)}{\sqrt{3\pi G}\, m_1t},\\
& H(t) \approx \frac{2}{3 t} + \frac{\sin (2 m_1 t + 2\alpha_1)}{3m_1 t^2},
\label{lateH}
\end{align}
where $\alpha_1$ is a phase factor depending on the values of $m_n$ and $\varphi_n(0)$.

\section{Scalar Perturbation and Sasaki-Mukhanov Variable $Q_n$}

In this section, we briefly review the scalar perturbation with multiple matter scalar fields,
based on Refs.~\cite{Hwang:1996gz,Taruya:1997iv,Gordon:2000hv},
and apply for the case of $N$ Lee-Wick scalar fields.
The metric perturbation is decomposed into
four scalar, four vector, and two tensor components.
For the scalar perturbation with $N$ matter scalar fields,
the total number of perturbation modes is $4+N$.
Imposing two constraints originating from (00)- and (0$i$)-components
of the Einstein's equation
and fixing two gauge degrees of freedom from the ($4+N$)-modes,
we have only $N$ dynamical degrees of freedom in the system.

Let us consider the linear scalar perturbation of the Friedmann-Robertson-Walker metric,
\begin{align}\label{pertFRW}
ds^2 =  -(1+ 2 A) dt^2 + 2a\partial_i B dx^i dt + a^2\left[
(1-2\psi)\delta_{ij} + 2\partial_i\partial_j E \right] dx^idx^i,
\end{align}
where $A$, $B$, $\psi$, and $E$ are four scalar modes.
Then the (00)- and (0$i$)-components of the perturbed Einstein's equation give the relations, respectively,
\begin{align}
&3H(\dot\psi+HA) + \frac{k^2}{a^2}\left[\psi + H(a^2\dot E- aB)\right] =
-4\pi G\delta\rho, \nn
\\
& \dot\psi + HA = -4\pi G\delta q, \label{delG0i}
\end{align}
where $k$ stands for the comoving wave number, and
\begin{align}
\delta\rho &= \sum_{n=1}^{N}(-1)^{n+1}\left[\dot\varphi_n
(\dot{\delta\varphi}_n-\dot\varphi_n A)
+ m_n^2\varphi_n\delta\varphi_n\right],
\nn \\
\delta q &=- \sum_{n=1}^{N}(-1)^{n+1}\dot\varphi_n\delta\varphi_n,
\end{align}
and $\delta\varphi_n$ represents the perturbed mode of the background scalar field $\varphi_n$.
The field equation for the matter fields is given by
\begin{align}\label{delvp}
\ddot{\delta\varphi}_n +3H\dot{\delta\varphi}_n + \frac{k^2}{a^2} \delta\varphi_n
+m_n^2\delta\varphi_n= -2m_n^2\varphi_n A + \dot\varphi_n\left[
\dot A + 3\dot\psi + \frac{k^2}{a^2}\left(a^2\dot E- aB\right)\right].
\end{align}
After fixing two gauge modes, we can completely determine the time evolution
of the scalar perturbation using the equations \eqref{delG0i} and \eqref{delvp}.
In other words, the resulting behaviors of the scalar perturbation satisfy
the remaining ($i$,$j$)-components of the perturbed Einstein's equation.

In order to fix two gauge degrees of freedom,
we introduce several gauge invariant quantities in terms of the scalar modes in the metric \eqref{pertFRW},
\begin{align}\label{defQn}
\Phi &\equiv A + aH(B-a\dot E) + a(B-a\dot E)^{\centerdot},
\nonumber \\
\Psi &\equiv \psi - aH(B-a\dot E).
\end{align}
We also introduce a gauge invariant definition,
which is refereed as Sasaki-Mukhanov variables~\cite{Sasaki:1986hm},
\begin{align}\label{SM}
Q_n &\equiv \delta\varphi_n + \frac{\dot\varphi_n}{H}\psi.
\end{align}
In this paper we choose the spatially flat gauge,
\begin{align}\label{flatgauge}
\psi=0,\qquad E=0.
\end{align}
Then we can replace gauge dependent variables  in \eqref{delG0i} and \eqref{delvp}
with the gauge invariant ones,
\begin{align}\label{flatgauge-2}
&\delta\varphi_n = Q_n,
\nonumber \\
&a^2\dot E - a B = \frac{\Psi}{H}.
\end{align}
Using these relations we can rewrite the constraints in \eqref{delG0i} as
\begin{align}\label{G0i-2}
\frac{k^2}{a^2}\Psi &= -4\pi G(\delta\rho - 3H\delta q)
\nonumber \\
&= -4\pi G\sum_{n=1}^{N}(-1)^{n+1}\left[\dot\varphi_n(\dot Q_n - \dot\varphi_n A)
-\ddot\varphi_n Q_n\right],
\nonumber \\
A&= -\frac{4\pi G}{H} \delta q = \frac{4\pi G}{H}\sum_{n=1}^{N}(-1)^{n+1}
\dot\varphi_n Q_n.
\end{align}
Plugging \eqref{flatgauge-2} and \eqref{G0i-2} into \eqref{delvp},
we express the right-hand side of \eqref{delvp} with the background variables and $Q_n$,
\begin{align}
{\rm RHS} &= -2m_n^2\varphi_n A + \dot\varphi_n\left(\dot A + \frac{k^2}{a^2}
\frac{\Psi}{H}\right)
\nonumber \\
&= 8\pi G\sum_{l=1}^N(-1)^{l+1}\Big[\frac1H \ddot\varphi_n\dot\varphi_lQ_l
+ 3\dot\varphi_n\dot\varphi_l Q_l + \frac{d}{dt}\left(\frac1H\right)\dot\varphi_n
\dot\varphi_l Q_l + \frac1H\dot\varphi_n\ddot\varphi_l Q_l\Big]
\nonumber \\
&=\frac{8\pi G}{a^3}
\sum_{l=1}^N(-1)^{l+1}\frac{d}{dt}\left(\frac{a^3}{H}
\dot\varphi_n\dot\varphi_l\right)Q_l,
\end{align}
where we used the background evolutions \eqref{bgHdot} and \eqref{bgmeq}.
We finally obtain $N$ equations for $Q_n$,
\begin{align}\label{maseq}
\ddot Q_n +3H\dot Q_n + \frac{k^2}{a^2} Q_n
+m_n^2Q_n -\frac{8\pi G}{a^3}
\sum_{l=1}^N(-1)^{l+1}\frac{d}{dt}\left(\frac{a^3}{H}
\dot\varphi_n\dot\varphi_l\right)Q_l=0.
\end{align}
With the Sasaki-Mukhanov variables $Q_n$,
the {\it comoving curvature} ${\cal R}$ is defined as
\begin{align}\label{Rgen}
{\cal R} = \psi-\frac{H}{\rho + p}\,\delta q
=H\times\left[
\frac{\sum_{n=1}^{N}(-1)^{n+1}\dot\varphi_nQ_n}{\sum_{m=1}^{N}(-1)^{m+1}\dot\varphi_m^2}\right].
\end{align}

\section{Evaluation of $Q_n$}\label{EvalQ}
In this section, we solve numerically \eqref{maseq} for $N=2$ to obtain $Q_n(t)$.
The background fields $\varphi_n(t)$ and $H(t)$ were already solved numerically as in Sec.~2.
In order to solve $Q_n(t)$ in this background,
first we perform the series expansion about the bouncing point $t=0$.
The background fields can be expanded as
\begin{align}\label{sphi}
\varphi_n(t)&= p_{n0} + p_{n1}t + p_{n2} t^2 + p_{n3} t^3 + \cdots, \\
H(t) &= h_0 + h_1 t + h_2 t^2 + h_3 t^3 + \cdots.\label{sH}
\end{align}
Plugging these expansions in the background field equations \eqref{bgHdot} and \eqref{bgmeq},
we can obtain the coefficients $p_{ni}$'s and $h_i$'s.
As we mentioned in Sec.~2, $\varphi_n$'s are even functions and $H$ is an odd function.
Therefore, $p_{ni}(i={\rm odd})=0$ and $h_i(i={\rm even})=0$.
In addition, applying the {\it symmetric} and {\it bouncing} conditions at $t=0$,
all the other coefficients are expressed in terms of $p_{n0}$.
[Note that $p_{n0} = \varphi_n(0)$. For the $N=2$ case,
only one of $p_{n0}$ is free since $m_1^2\varphi_1^2(0) = m_2^2\varphi_2^2(0)$
from  \eqref{bgHsq}.]
% with $H(0)=0$ and $\dot\varphi_n(0)=0$.]
The coefficients are then determined as follows;
\begin{align}
p_{n2} &= -\frac{m_n^2}{2}\,p_{n0},
\nonumber \\
p_{n4} &= \frac{m_n^4}{24}p_{n0},
\label{vpseries} \\
p_{n6} &=-\frac{m_n^6}{30}\left[4\pi G \sum_{l=1}^{N} (-1)^{l+1} \frac{m_l^4}{m_n^4} p_{l0}^2 + \frac{1}{24}\right]p_{n0},
\nonumber \\
&\cdots, \nonumber \\ \nonumber \\
h_1 &= 0,
\nonumber \\
h_3&= -\frac{4\pi G}{3}\sum_{n=1}^{N}(-1)^{n+1} m_n^4 p_{n0}^2,
\label{Hseries2} \\
h_5&=\frac{4\pi G}{15}\sum_{n=1}^{N} (-1)^{n+1}m_n^6 p_{n0}^2,
\nonumber \\
&\cdots. \nonumber
\end{align}
The corresponding behavior of the scale factor near the bouncing point is given by
\begin{align}
a(t) = 1 + \frac14 h_3 t^4 + \frac16 h_5 t^6 + \cdots
\label{aaa}
\end{align}
with the normalization $a(0)=1$.
As it was pointed out in Sec.~2.1, for the bouncing from contraction to expansion,
one should have $\ddddot{a}(0) >0$.
This is satisfied when $h_3>0$ which implies $|m_2|>|m_1|$ using \eqref{Hseries2}.

Now we expand $Q_n$ in series as
\begin{align}\label{Qtgen=0}
Q_n(t) &= t^s (q_{n0} + q_{n1} t
+q_{n2} t^2 + q_{n3} t^3 + q_{n4} t^4 + \cdots).
\end{align}
Since the comoving curvature ${\cal R}$ in  \eqref{Rgen} can be regular
even when $Q_n$ is singular, we introduced $t^s$ factor in the front
(and it can be $s<0$).
Plugging the expansions \eqref{vpseries}, \eqref{Hseries2}, and \eqref{Qtgen=0}
into the $Q$-equation~\eqref{maseq},
we obtain two linearly independent solutions for $Q_n$.
The one is an even function with $s=-2$ and $q_{ni}(i={\rm odd})=0$,
and the other is an odd function with $s=0$ and $q_{ni}(i={\rm even})=0$;
\begin{align}
Q^{\rm even}_n(t) &= t^{-2} (q_{n0} +q_{n2} t^2  + q_{n4} t^4 + \cdots),\\
Q^{\rm odd}_n(t) &= q_{n1} t + q_{n3} t^3 + q_{n5} t^5 +  \cdots.
\end{align}
In addition, we are also provided with the relations for $q_{ni}$;
for the even case,
\begin{align}
q_{20} &= \frac{m_2}{m_1}\; q_{10}, \label{q1}\\
q_{22} &= \frac{m_1}{m_2}\; q_{12} -\frac{(m_1^2-m_2^2)(5k^2+m_1^2+m_2^2)}{30m_1m_2}\;q_{10},
\end{align}
and for the odd case,
\begin{align}
q_{21} &= \frac{m_1}{m_2}\; q_{11}, \\
q_{23} &= \frac{m_2}{m_1}\; q_{13} -\frac{(m_1^2-m_2^2)k^2}{6m_1m_2}\;q_{11}. \label{q4}
\end{align}
All the other $q_{ni}$'s can also be determined
by the background parameters ($m_1$, $m_2$) and the fluctuation parameters ($k$, $q_{10}$, $q_{12}$, $q_{11}$, $q_{13}$).
For a given comoving wave number $k$, $q_{10}$ and $q_{12}$ ($q_{11}$ and $q_{13}$)
are free parameters for the even (odd) case.
These free parameters are the {\it shooting parameters} in numerical calculations for $Q_n$.
Since the $Q$-equation is linear, constant$\times Q_n$ is also a solution.
One of the shooting parameters plays a role of constant multiplication,
and the other plays the role of real shooting parameter.
From numerical calculations, we observe that the shape of the solutions
does not depend on the value of the second parameter, which means that
the solutions are attractive.

The numerical solutions of $Q_n$ for the even and the odd cases are
plotted in Fig.~2. As the background scalar fields $\varphi_n$'s enter the oscillating regime,
the universe settles to the matter-dominated expansion.
Then the perturbation $Q_n$ oscillates about zero apparently with a constant amplitude
in the same frequency with the background fields.
The amplitude of $Q_1$ is much larger than that of $Q_2$,
so the perturbation properties such as the comoving curvature ${\cal R}$
are mainly determined by $Q_1$.

For future purpose, let us analyze the late-time behavior of the perturbation.
Since $Q_1$ is dominant, neglecting $Q_2$ in \eqref{maseq},
the perturbation equation for $Q_1$ is approximated by
\begin{align}\label{Q1late}
\ddot Q_1 +3H\dot Q_1 + \left(\frac{k^2}{a^2}+ m_1^2\right) Q_1
-\frac{8\pi G}{a^3}\frac{d}{dt}\left(\frac{a^3}{H}
\dot\varphi_1^2 \right)Q_1 \approx 0.
\end{align}
Plugging the asymptotic solutions for $\varphi_1$ and $H$ presented in
\eqref{latephi} and \eqref{lateH}, and $a(t)\approx \alpha t^{2/3}$,
and selecting the dominant terms at late times  ($m_1t\gg 1$),
the above equation becomes
\begin{align} \label{lateQQ}
\ddot Q_1 + \frac{2}{t}\,\dot Q_1 +\left[\frac{k^2}{\alpha^2 t^{4/3}}
+ m_1^2-\frac{4m_1}{t}\sin (2m_1 t + 2\alpha_1) \right] Q_1\approx 0.
\end{align}
%%%%%%%%%%%%%%%%%%%%%%%%%%%%%%%%%%%%%%%%%%%%%%%%%%%%%%%%%%%%%%%
\begin{figure}%[!h]
\centerline{\epsfig{figure=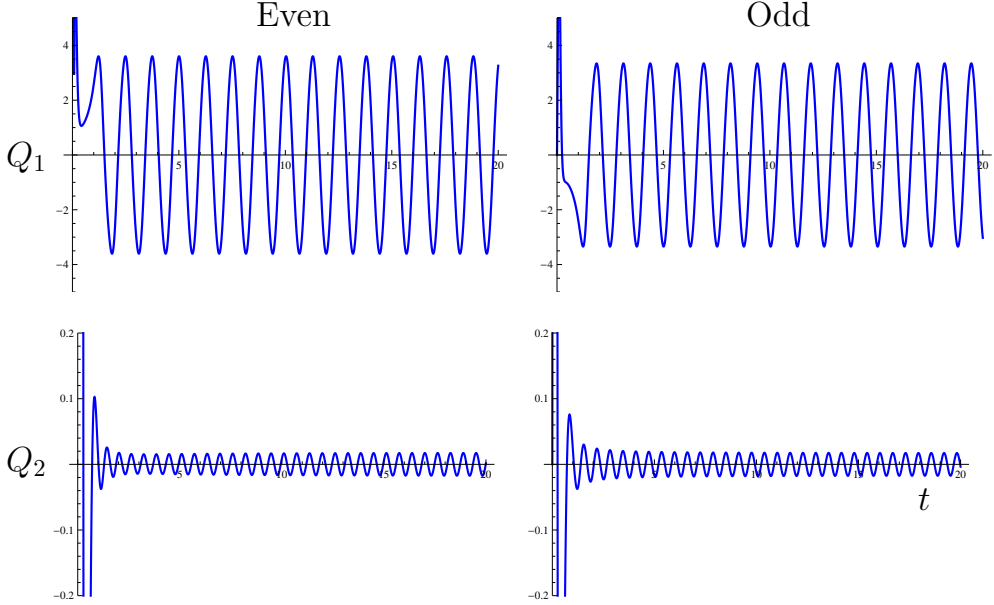,height=80mm}}
\caption{
\small Plot of the Sasaki-Mukhanov variable $Q_n$
for the even case ($q_{10} = 10^{-21}$, $q_{12} = 10$)
and for the odd case ($q_{11} = 10^{2}$, $q_{13} = 10^5$).
$Q_n$ settles down to an oscillation with an almost constant amplitude.
$Q_1$ has a much larger amplitude.
The $t<0$ region is evenly and oddly symmetric.
For the even case, $Q_n$ diverges at $t=0$,
but is not visible in the figure.
}
\label{FIG2}
\end{figure}
%%%%%%%%%%%%%%%%%%%%%%%%%%%%%%%%%%%%%%%%%%%%%%%%%%%%%%%%%%%%%%%
Among five terms in the equation, the first term and the fourth (mass) term are the most dominant
and comparable in magnitude. These terms give a constant oscillation.
The next dominant terms are the second (friction) term and
the last (sinusoidal) term which are comparable in magnitude.
These two terms add $t^{-1}$ damped oscillation behavior in $Q_1$.
The third ($k$-) term is comparable with these two terms in magnitude
during the {\it intermediate} period for {\it large} $k$.
In this period, the damped oscillation in $Q_1$ behaves as $t^{-1/3}$
due to the $k$-term.
The numerical results shown in Fig.~2 contain the constant and the damped oscillations in principle,
although the damped oscillation is not visible because it is tiny compared with the constant oscillation.
These two oscillating patterns play an important role in the comoving curvature ${\cal R}$
as we shall see in the next section.

\section{Initial Perturbation and Power Spectrum}
\subsection{Comoving curvature ${\cal R}$}
For $N=2$, the comoving curvature ${\cal R}$ in  \eqref{Rgen} becomes
\begin{align}\label{RR}
{\cal R} = \frac{H}{\dot\varphi_1^2-\dot\varphi_2^2}
\left(\dot\varphi_1 Q_1 - \dot\varphi_2 Q_2 \right)
\equiv f_1 Q_1 - f_2 Q_2.
\end{align}
%The main contribution in ${\cal R}$ is from $Q_1$.
The factor $f_n$ is a function of the background fields,
and is an even function.
Therefore, the evenness/oddness of ${\cal R}$
is the same with that of $Q_n$.
The factor function $f_n$ becomes a periodic function at late times. (See Fig.~3.)
Whenever $\dot\varphi_1^2=\dot\varphi_2^2$, $f_n$ diverges.
However, this divergence does not make the evaluation of ${\cal R}$
unphysical in the whole time~\cite{Allen:2004vz,Kim:2006ju}.
One may ignore simply the diverging period only
while keeping the rest as physical.
%%%%%%%%%%%%%%%%%%%%%%%%%%%%%%%%%%%%%%%%%%%%%%%%%%%%%%%%%%%%%%%
\begin{figure}%[!h]
\centerline{\epsfig{figure=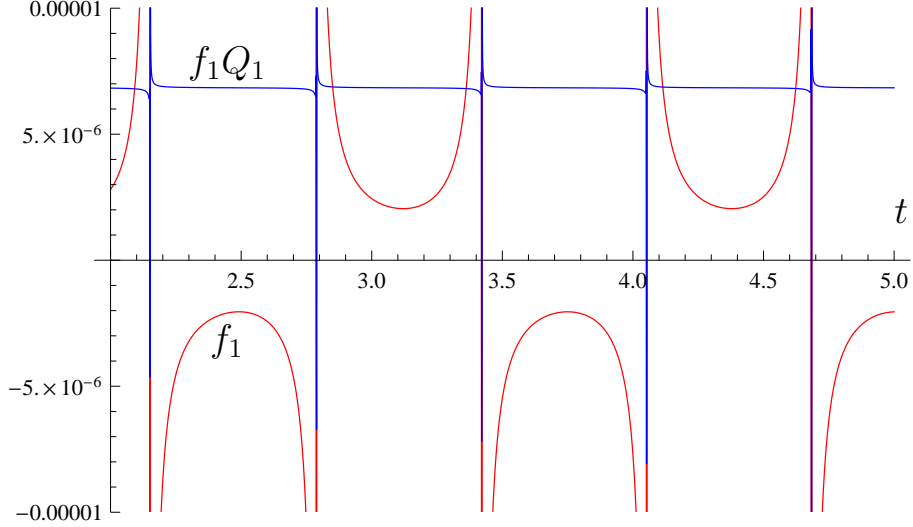,height=70mm}}
\caption{
\small Plot of the factor function $f_1$ and $f_1Q_1$ in ${\cal R}$.
(Only the odd case is plotted, but the even case is similar.)
The shape of $f_2Q_2$ is also similar, but the overall level is lower
and the period is half.
}
\label{FIG3}
\end{figure}
%%%%%%%%%%%%%%%%%%%%%%%%%%%%%%%%%%%%%%%%%%%%%%%%%%%%%%%%%%%%%%%

As we observed in the previous section,
$Q_n$ has a periodic oscillating behavior.
The frequency is the same with that of $f_n$, but the phase is a little bit off.
Then the shape of $f_nQ_n$ becomes as shown in Fig.~3.
Considering one period, except for the vicinity of the divergence,
$f_nQ_n$ stays almost constant.
In the successive periods, this pattern is maintained, and
the value of $f_nQ_n$ does not seem to change
due to the apparently constant amplitude of $Q_n$.
However, as it was discussed in the previous section,
there is a damped oscillation in $Q_n$ in addition to the constant oscillation.
The effect of this damped oscillation will be implied in the pattern of $f_nQ_n$.
Therefore, as periods proceed, the value of $f_nQ_n$ will decrease slightly.
%%%%%%%%%%%%%%%%%%%%%%%%%%%%%%%%%%%%%%%%%%%%%%%%%%%%%%%%%%%%%%%
\begin{figure}%[!h]
\centerline{\epsfig{figure=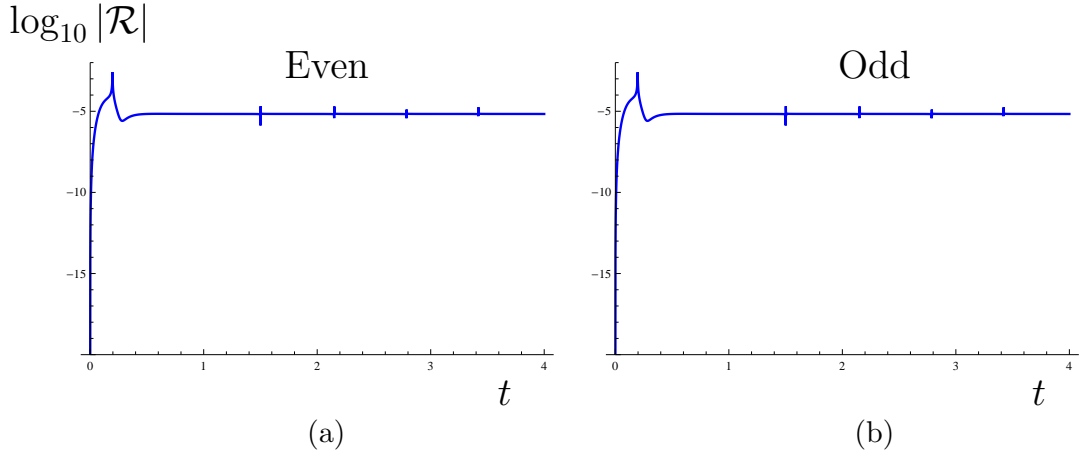,height=60mm}}
\caption{
\small Plot of the comoving curvature ${\cal R}$.
The spikes indicate divergences.
Except for the divergent region,
the curvature approaches an apparently constant value soon.
The constant value can be adjusted by varying the values of
shooting parameters.
The decaying in ${\cal R}$ from the damped oscillation in $Q_1$
is inherent, but not visible.
}
\label{FIG4}
\end{figure}
%%%%%%%%%%%%%%%%%%%%%%%%%%%%%%%%%%%%%%%%%%%%%%%%%%%%%%%%%%%%%%%
%%%%%%%%%%%%%%%%%%%%%%%%%%%%%%%%%%%%%%%%%%%%%%%%%%%%%%%%%%%%%%%
\begin{figure}%[!h]
\centerline{\epsfig{figure=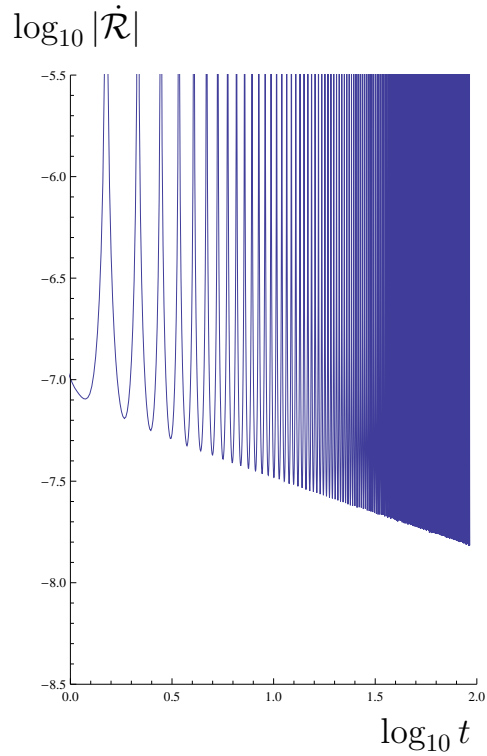,height=100mm}}
\caption{
\small Plot of $\log_{10}|\dot{\cal R}|$ vs. $\log_{10}t$.
The slope indicates that ${\cal R}$ contains the decaying mode.
(We plotted only for the odd case, but the even case is similar.)
From the slope, the decaying is proportional to $t^{-1/3}$.
The $t^{-1}$ decaying behavior is not achieved within the numerical domain.
}
\label{FIG5}
\end{figure}
%%%%%%%%%%%%%%%%%%%%%%%%%%%%%%%%%%%%%%%%%%%%%%%%%%%%%%%%%%%%%%%

The behavior of $f_nQ_n$ is embodied in the comoving curvature ${\cal R}$ from \eqref{RR}.
As it is shown in Fig.~4, ${\cal R}$ stays constant except the vicinity
of the divergent points,
while the effect of the damped oscillation is not very visible.

What is remaining is how to extract the damped oscillation from our numerical
results of ${\cal R}$.
In order to see this,
we plot $\log_{10}|\dot{\cal R}|$ vs. $\log_{10}t$ in Fig.~5.
The slope is $\approx -1/3$, which indicates
that the decaying hidden in constant ${\cal R}$
behaves as ${\cal R}^{\rm decaying} \propto t^{-1/3}$.
This decaying behavior results from the damped oscillation in $Q_1$ controlled
by the $k$-term in  \eqref{lateQQ} discussed in the previous section.
In the numerical domain shown in the plot, the effect of the $k$-term is
considerable. If the numerical calculation could go further in time,
one would be able to see the decaying behavior transit from
$t^{-1/3}$ to $t^{-1}$.
Unfortunately, however, the numerical integration stops before the
transition due to numerical error accumulated.

\subsection{Initial Perturbation}
Since the comoving curvature ${\cal R}$ has been obtained,
let us discuss the initial perturbation considering the modes of ${\cal R}$.
In the previous section, we obtained two linearly independent modes
of ${\cal R}$, the even ($E$-) and the odd ($O$-) modes.
In general during the expanding phase ($t>0$),
the curvature consists of the constant ($C$-) and the decaying ($D$-) modes
which are linearly independent.
The $C$- and $D$-modes can be obtained from a linear combination of the $E$- and $O$-modes.
These $C$- and $D$-modes are those discussed in the previous subsection.

During the contracting phase ($t<0$),
the decaying mode becomes a growing ($G$-) mode due to symmetry.
The $G$-mode provides the initial perturbation in the bouncing universe.
The initial perturbation is produced in the contracting phase
deep inside the horizon, and then grows.
The perturbation produced in this way is frozen
after it crosses the horizon in the expanding phase ($t>0$),
and should produce the observed value in the power spectrum.

In order to discuss the initial quantum fluctuation and the vacuum state,
let us introduce the conformal time $\eta$ and a new field $v$ for canonical quantization
by~\footnote{We fix the bouncing point as $\eta(t=0)=0$.}
\begin{align}
dt=a d\eta,\qquad
v=aQ_1.
\end{align}
We consider only the contribution of $Q_1$ which is dominant at late times
as discussed in Sec.~\ref{EvalQ}.
The functional behavior of $v$ at early times ($\eta\ll 0$) can be read
from that at late times ($\eta\gg 0$), due to the symmetric property of our setting.
The action for the new field is written by
\begin{align}
S= \int d\eta dx^3 \left[ \frac{1}{2} (\partial_\eta \tilde{v})^2
- \frac{1}{2} (\partial_i \tilde{v})^2 +\frac{1}{2}\frac{z''}{z}\tilde{v}^2 \right],
\end{align}
where $'\equiv d/d\eta$, $z=a\dot\varphi/H$.
The general solution of the equation of motion is given by
\begin{align}
\tilde{v}(\eta,\vec{x})  = \int \frac{d^3k}{(2\pi)^{3/2}}
\left[ v(\eta;k) a_{\vec{k}}  +
v^* (\eta;k) a^\dag_{-\vec{k}} \right] e^{i\vec{k}\cdot\vec{x}},
\end{align}
where $v(\eta;k)$ satisfies
\begin{align}\label{veq}
v'' + \left(k^2 - \frac{z''}{z}\right) v = 0,
\end{align}
and is normalized as
\begin{align}\label{vnorm}
v v^{*\prime} - v^* v' = i.
\end{align}

At early times, $\eta \ll 0$, the field equation \eqref{veq} can be approximated by
\begin{align}\label{vlate}
v''(\eta) + \left[k^2 -\frac2{\eta^2} + \frac{m_1^2\alpha^6}{81} \eta^4
-\frac{4 m_1 \alpha^3}{3}\eta\,\sin \left(\frac{2m_1\alpha^3}{27} \eta^3
+ 2 \alpha_1\right)\right] v(\eta) \approx 0,
\end{align}
where we used $a\approx \alpha t^{2/3} = \alpha^3\eta^2/9$ and $t=\alpha^3\eta^3/27$.
This equation can be obtained from  \eqref{veq},
or from a direct transformation of \eqref{lateQQ}.
The horizon crossing for a given wavelength occurs when
\begin{align}
\lambda_{ph} = \frac{a}{k} \sim  \left| H^{-1} \right|
\quad\Longrightarrow\quad
|k\eta| \sim 2.
\end{align}
In the far subhorizon limit, $|k\eta| \gg 2$,
the equation \eqref{vlate} can be further approximated by
\begin{align}
\frac{d^2v(\eta)}{d(k\eta)^2}  + \frac{m_1^2\alpha^6}{81k^6}(k\eta)^4 v(\eta) \approx 0.
\end{align}
The solution to this equation satisfying the normalization condition \eqref{vnorm} is given by
\begin{align}\label{vHankel}
v(\eta) = \sqrt{\frac{\pi\eta}{12}}
\left[ A_1 H^{(1)}_{\frac{1}{6}} \left(\frac{m_1\alpha^3}{27} \eta^3\right)
+ A_2 H^{(2)}_{\frac{1}{6}} \left(\frac{m_1\alpha^3}{27} \eta^3\right) \right],
\end{align}
where $|A_2|^2 - |A_1|^2 = 1$ and $H_n^{(1,2)}$ represent the Hankel functions.
We adopt the Bunch-Davies vacuum for the production of the
initial perturbation at $\eta=\eta_i \ll 0$
by taking only the positive energy mode, $A_1=0$.
Then the vacuum solution  \eqref{vHankel} at $\eta\ll 0$ becomes
\begin{align}\label{v_grow}
v(\eta) \approx \sqrt{\frac{9}{2m_1\alpha^3}} \;\frac{1}{\eta}\;
\exp{\left[-i\left(\frac{m_1\alpha^3}{27}\eta^3 - \frac{\pi}{3}\right)\right]}.
\end{align}
This solution represents the initial perturbation produced
{\it deep inside the horizon},
and its amplitude grows in time.
It can be transformed to the original variables at $t\ll 0$,
\begin{align}\label{Q_grow}
Q_1(t) = \frac{v}{a} \propto   \frac{1}{t} \exp \left(-im_1t\right).
\end{align}
During the expanding phase ($t>0$), this solution represents
the $t^{-1}$ damped oscillation analyzed in the previous section,
which gives  ${\cal R}^{\rm decaying} \propto t^{-1}$.

The functional dependence on $\eta$ of the initially growing perturbation \eqref{v_grow}
is somewhat different from the initial perturbation
studied in inflation or in power-law expansion in the literature.
In those, $v \approx 1/\sqrt{2k}\; \exp (- ik\eta)$.
The difference originates from the {\it oscillating behavior} of the
background fields in the Lee-Wick model,
which provides the $\eta^4$-term in \eqref{vlate} as the most dominant term.

\subsection{Analysis of ${\cal R}$ modes and late-time power spectrum}
In this subsection, let us discuss the modes of ${\cal R}$.
The initial growing mode of ${\cal R}$ can be obtained from
the results of the previous section.
Since $Q_1$ has the dominant contribution to ${\cal R}$,
the $G$-mode of ${\cal R}$ at $t\ll 0$ is obtained from  \eqref{Q_grow},
\begin{align}
{\cal R}^{\rm growing} \approx f_1Q_1^{\rm growing} \propto \frac{1}{t}.
\end{align}
This $G$-mode can be extracted from the $E$- and $O$-modes by a linear combination.
Each of the even and the odd mode contains the growing and the constant modes in itself.
Then, the growing mode can be obtained as following;
\begin{align}
{\cal R}(t \ll0)
&= c_1 {\cal R}^{\rm even} + c_2 {\cal R}^{\rm odd} \\
&= c_1 \left[{\cal R}^{\rm even-growing} + {\cal R}^{\rm even-const}\right]
+ c_2 \left[{\cal R}^{\rm odd-growing} + {\cal R}^{\rm odd-const}\right] \\
&= c_1 {\cal R}^{\rm even-growing} + c_2 {\cal R}^{\rm odd-growing} \\
&\equiv {\cal R}^{\rm growing},
\end{align}
where $c_1$ and $c_2$ are fixed such that
$c_1 {\cal R}^{\rm even-const} + c_2{\cal R}^{\rm odd-const} =0$.
Then the constant modes in ${\cal R}$ are canceled each other, and
only the growing mode remains and plays the role of initial perturbation.

Once one fixes $c_1$ and $c_2$ as above, the linear combination of
${\cal R}^{\rm even}$ and ${\cal R}^{\rm odd}$ is fixed.
At $t\gg 0$, the comoving curvature ${\cal R}$ as a result of this linear combination
becomes
\begin{align}
{\cal R}(t\gg 0)
&= c_1 {\cal R}^{\rm even} + c_2 {\cal R}^{\rm odd} \\
&= c_1 \left[{\cal R}^{\rm even-decaying} + {\cal R}^{\rm even-const}\right]
+ c_2 \left[{\cal R}^{\rm odd-decaying} + {\cal R}^{\rm odd-const}\right] \\
&= \left[c_1 {\cal R}^{\rm even-decaying}
+c_2 {\cal R}^{\rm odd-decaying}\right] + \left[c_1 {\cal R}^{\rm even-const} + c_2 {\cal R}^{\rm odd-const}\right] \\
&\equiv {\cal R}^{\rm decaying}  +{\cal R}^{\rm const} \quad\approx {\cal R}^{\rm const} .
\end{align}
As it can be seen in the schematic picture in Fig.~6,
the constant modes in ${\cal R}^{\rm even}$ and ${\cal R}^{\rm odd}$
do not cancel each other because the odd mode changes its signature at $t>0$.
Therefore, at late times during the expanding phase,
the comoving curvature approaches a constant.
This constant must reproduce the observed value of the power spectrum,
\begin{align}
{\cal P}_{\cal R} = \frac{k^3}{2\pi^2}|{\cal R}|^2 \approx 10^{-9}.
\label{PS}
\end{align}

%%%%%%%%%%%%%%%%%%%%%%%%%%%%%%%%%%%%%%%%%%%%%%%%%%%%%%%%%%%%%%%
\begin{figure}%[!h]
\centerline{\epsfig{figure=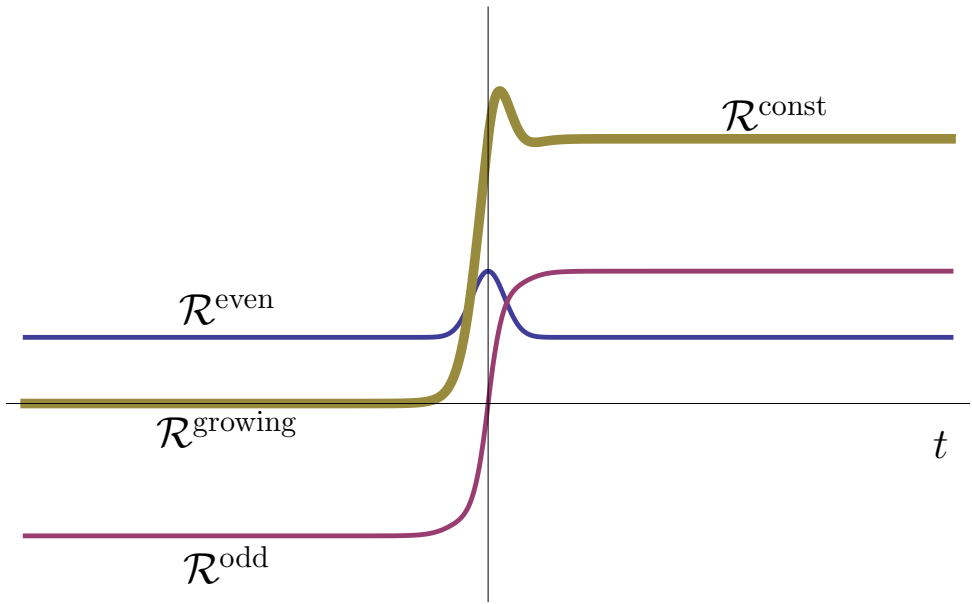,height=80mm}}
\caption{
\small Schematic picture of the linear combination of ${\cal R}^{\rm even}$
and ${\cal R}^{\rm odd}$.
At $t\ll 0$, the initial growing mode ${\cal R}^{\rm growing}$ can be extracted
from the combination.
At $t\gg 0$, this combination provides the constant mode ${\cal R}^{\rm const}$
which is responsible for the late-time power spectrum.
}
\label{FIG6}
\end{figure}
%%%%%%%%%%%%%%%%%%%%%%%%%%%%%%%%%%%%%%%%%%%%%%%%%%%%%%%%%%%%%%%
Although the schematic story is as above, it will be very difficult to extract
the appropriate ${\cal R}^{\rm growing}$ mode from numerical calculations.
First of all, our numerical results could not reach the region of large enough $|t|$
where the $t^{-1}$ decaying mode is realized.
Even when the numerical technique is improved to reach such a large $|t|$ region,
combining the even and the odd modes in such a way as to meet the both of
the initial condition \eqref{v_grow} and the late-time power spectrum \eqref{PS}
simultaneously, is nontrivial.
Since $Q_n$ has an oscillating motion, one needs to match the amplitude
as well as the phase exactly by adjusting the initial shooting parameters.
It is not clear whether or not the numerical error will admit such an exact matching.

\section{Conclusions}
In this paper, we investigated the scalar perturbation
of the Lee-Wick bouncing universe driven by two scalar fields,
one of which is an ordinary field and the other
is a ghost field.
The universe contracts initially and then expands, owing to the
energy-condition violating ghost field.
For the analytical purpose, we considered a symmetric evolution
about the bouncing point $t=0$;
The scale factor $a$ and the scalar fields $\varphi_n$'s are even functions.
Near the bouncing point, the universe undergoes an accelerating
contraction and expansion.
If we set the ghost field $\varphi_2$ is more massive than
the ordinary field $\varphi_1$, $m_2>m_1$,
the universe is mainly driven by the ordinary field at large  $|t|$.
The ordinary field $\varphi_1$ at large $t$ exhibits a damped oscillation
and plays the role of the pressureless dust driving
the universe to a matter-dominated expansion.
At $t<0$, the evolution is opposite.

We investigated the scalar perturbation in this background.
For the scalar perturbation,
we adopted the spatially flat gauge and numerically solved
the gauge invariant Sasaki-Mukhanov variable $Q_n$.
In calculating $Q_n$, we performed a series expansion about $t=0$,
and found that there exist two linearly independent solutions.
One is an {\it even} function and the other is an {\it odd} function.
This kind of approach in calculating $Q_n$ about $t=0$
reduced difficulties in dealing with the bouncing point.
(If one starts to solve numerically $Q_n$ from $t<0$,
it is very difficult to overcome the bouncing point.)
The resulting evenness/oddnes of $Q_n$ coming from the symmetry of the background
saved the labor for the half of calculation, and
enabled us to analyze the corresponding physics in an easy way.

From the numerical results of $Q_n$,
we evaluated the comoving curvature ${\cal R}$.
From the linear combination of the even and the odd modes of ${\cal R}$,
we showed that the growing mode and the constant modes can be extracted in principle.
The growing mode is responsible for the initial scalar perturbation
produced deep inside the horizon during the contracting phase.
We found that the initial perturbation in the Bunch-Davies vacuum has
a form \eqref{v_grow} which is different from those known in the literature.
This difference comes from the oscillating nature of the background fields.
This growing mode is connected to the constant mode
achieved after the perturbation crosses out the horizon
during the expanding phase.
This constant mode explains the late-time power spectrum.

Due to the numerical difficulties,
we could not see the $t^{-1}$ decaying/growing
behavior in ${\cal R}$. With a better numerical technique,
we wish to come back to achieve the complete decaying behavior in the future.
We considered only for the $N=2$ Lee-Wick model,
but it will also be interesting to study for the $N>2$ cases.

\subsection*{Acknowledgements}
We are grateful to Jinn-Ouk Gong, Seoktae Koh, Seokcheon Lee, Takahiro Tanaka,
and in particular, Jai-chan Hwang for very helpful discussions.
This work was supported by the Korea Research Foundation (KRF) grant
funded by the Korea government (MEST) No. 2009-0070303 (I.Y.),
and No. 2011-0009972 (O.K.).


\begin{thebibliography}{99}

%\cite{Lee:1969fy}
\bibitem{Lee:1969fy}
  T.~D.~Lee and G.~C.~Wick,
  ``Negative Metric and the Unitarity of the S Matrix,''
  Nucl.\ Phys.\  B {\bf 9}, 209 (1969);
  %%CITATION = NUPHA,B9,209;%%
%\cite{Lee:1970iw}
%\bibitem{Lee:1970iw}
%  T.~D.~Lee and G.~C.~Wick,
  ``Finite Theory of Quantum Electrodynamics,''
  Phys.\ Rev.\  D {\bf 2}, 1033 (1970).
  %%CITATION = PHRVA,D2,1033;%%

%\cite{Grinstein:2007mp}
\bibitem{Grinstein:2007mp}
  B.~Grinstein, D.~O'Connell and M.~B.~Wise,
  ``The Lee-Wick standard model,''
  Phys.\ Rev.\  D {\bf 77}, 025012 (2008)
  [arXiv:0704.1845 [hep-ph]].
  %%CITATION = PHRVA,D77,025012;%%

%\cite{Grinstein:2007iz}
\bibitem{Grinstein:2007iz}
  B.~Grinstein, D.~O'Connell and M.~B.~Wise,
  ``Massive Vector Scattering in Lee-Wick Gauge Theory,''
  Phys.\ Rev.\  D {\bf 77}, 065010 (2008)
  [arXiv:0710.5528 [hep-ph]].
  %%CITATION = PHRVA,D77,065010;%%


%\cite{Grinstein:2008bg}
\bibitem{Grinstein:2008bg}
  B.~Grinstein, D.~O'Connell and M.~B.~Wise,
  ``Causality as an emergent macroscopic phenomenon: The Lee-Wick O(N) model,''
  Phys.\ Rev.\  D {\bf 79}, 105019 (2009)
  [arXiv:0805.2156 [hep-th]].
  %%CITATION = PHRVA,D79,105019;%%


%\cite{Carone:2008iw}
\bibitem{Carone:2008iw}
  C.~D.~Carone and R.~F.~Lebed,
  ``A Higher-Derivative Lee-Wick Standard Model,''
  JHEP {\bf 0901}, 043 (2009)
  [arXiv:0811.4150 [hep-ph]].
  %%CITATION = JHEPA,0901,043;%%

%\cite{Cho:2010hj}
\bibitem{Cho:2010hj}
  I.~Cho and O.~K.~Kwon,
  ``Generalized Lee-Wick Formulation from Higher Derivative Field Theories,''
  Phys.\ Rev.\  D {\bf 82}, 025013 (2010)
  [arXiv:1003.2716 [hep-th]].
  %%CITATION = PHRVA,D82,025013;%%

%\cite{Cai:2008qw}
\bibitem{Cai:2008qw}
  Y.~F.~Cai, T.~t.~Qiu, R.~Brandenberger and X.~m.~Zhang,
  ``A Nonsingular Cosmology with a Scale-Invariant Spectrum of Cosmological
  Perturbations from Lee-Wick Theory,''
  Phys.\ Rev.\  D {\bf 80}, 023511 (2009)
  [arXiv:0810.4677 [hep-th]].
  %%CITATION = PHRVA,D80,023511;%%

%\cite{Cai:2009fn}
\bibitem{Cai:2009fn}
  Y.~F.~Cai, W.~Xue, R.~Brandenberger and X.~Zhang,
  ``Non-Gaussianity in a Matter Bounce,''
  JCAP {\bf 0905}, 011 (2009)
  [arXiv:0903.0631 [astro-ph.CO]];
  %%CITATION = JCAPA,0905,011;%%

%\cite{Karouby:2010wt}
%\bibitem{Karouby:2010wt}
  J.~Karouby and R.~Brandenberger,
  ``A Radiation Bounce from the Lee-Wick Construction?,''
  Phys.\ Rev.\  D {\bf 82}, 063532 (2010)
  [arXiv:1004.4947 [hep-th]];
  %%CITATION = PHRVA,D82,063532;%%

%\cite{Karouby:2011wj}
%\bibitem{Karouby:2011wj}
  J.~Karouby, T.~Qiu and R.~Brandenberger,
  ``On the Instability of the Lee-Wick Bounce,''
  Phys.\ Rev.\  D {\bf 84}, 043505 (2011)
  [arXiv:1104.3193 [hep-th]];
  %%CITATION = PHRVA,D84,043505;%%

%\cite{Cai:2011ci}
%\bibitem{Cai:2011ci}
  Y.~F.~Cai, R.~Brandenberger and X.~Zhang,
  ``Preheating a bouncing universe,''
  Phys.\ Lett.\  B {\bf 703}, 25 (2011)
  [arXiv:1105.4286 [hep-th]].
  %%CITATION = PHLTA,B703,25;%%

 \bibitem{Zhang:2010bb}
  J.~Zhang, Z.~G.~Liu and Y.~S.~Piao,
  ``Amplification of curvature perturbations in cyclic cosmology,''
  Phys.\ Rev.\  D {\bf 82}, 123505 (2010)
  [arXiv:1007.2498 [hep-th]];
  %%CITATION = PHRVA,D82,123505;%%
  
 %\cite{Xue:2011nw}
%\bibitem{Xue:2011nw}
  B.~Xue and P.~J.~Steinhardt,
  ``Evolution of curvature and anisotropy near a nonsingular bounce,''
  arXiv:1106.1416 [hep-th];
  %%CITATION = ARXIV:1106.1416;%%
  
%\cite{Easson:2011zy}
%\bibitem{Easson:2011zy}
  D.~A.~Easson, I.~Sawicki, A.~Vikman,
  ``G-Bounce,''
  [arXiv:1109.1047 [hep-th]].


%\cite{Hwang:1996gz,Taruya:1997iv}
\bibitem{Hwang:1996gz}
  J.~c.~Hwang,
  ``Cosmological perturbations with multiple scalar fields,''
  arXiv:gr-qc/9608018.
  %%CITATION = GR-QC/9608018;%%

%\cite{Taruya:1997iv}
\bibitem{Taruya:1997iv}
  A.~Taruya and Y.~Nambu,
  ``Cosmological perturbation with two scalar fields in reheating after
  inflation,''
  Phys.\ Lett.\  B {\bf 428}, 37 (1998)
  [arXiv:gr-qc/9709035].
  %%CITATION = PHLTA,B428,37;%%

%\cite{Gordon:2000hv}
\bibitem{Gordon:2000hv}
  C.~Gordon, D.~Wands, B.~A.~Bassett and R.~Maartens,
  ``Adiabatic and entropy perturbations from inflation,''
  Phys.\ Rev.\  D {\bf 63}, 023506 (2001)
  [arXiv:astro-ph/0009131].
  %%CITATION = PHRVA,D63,023506;%%

%\cite{Sasaki:1986hm}
\bibitem{Sasaki:1986hm}
  M.~Sasaki,
  ``Large Scale Quantum Fluctuations in the Inflationary Universe,''
  Prog.\ Theor.\ Phys.\  {\bf 76}, 1036 (1986);
  %%CITATION = PTPKA,76,1036;%%

%\cite{Mukhanov:1988jd}
%\bibitem{Mukhanov:1988jd}
  V.~F.~Mukhanov,
  ``Quantum Theory of Gauge Invariant Cosmological Perturbations,''
  Sov.\ Phys.\ JETP {\bf 67} (1988) 1297
  [Zh.\ Eksp.\ Teor.\ Fiz.\  {\bf 94N7} (1988) 1].
  %%CITATION = ZETFA,94N7,1;%%

%\cite{Allen:2004vz}
\bibitem{Allen:2004vz}
  L.~E.~Allen and D.~Wands,
  ``Cosmological perturbations through a simple bounce,''
  Phys.\ Rev.\  D {\bf 70}, 063515 (2004)
  [arXiv:astro-ph/0404441].
  %%CITATION = PHRVA,D70,063515;%%

%\cite{Kim:2006ju}
\bibitem{Kim:2006ju}
  H.~S.~Kim and J.~c.~Hwang,
  ``Evolution of linear perturbations through a bouncing world model: Is the
  Harrison-Zel'dovich spectrum possible via bounce?,''
  Phys.\ Rev.\  D {\bf 75}, 043501 (2007)
  [arXiv:astro-ph/0607464].
  %%CITATION = PHRVA,D75,043501;%%



\end{thebibliography}
\end{document}